\documentclass{article}
\usepackage{spconf,amsmath,epsfig, hyperref}
\usepackage[utf8]{inputenc}
\usepackage{graphicx}
\usepackage{subfigure}
\usepackage{xcolor}
\usepackage{booktabs}
\usepackage{placeins}
\usepackage{hyperref}
\usepackage{comment}
\usepackage{cite}
\usepackage{balance}

\usepackage{url}
\title{QSpeckleFilter: a quantum machine learning approach\\
for SAR speckle filtering}

\name{\begin{tabular}{c}Francesco Mauro$^{a,1}$, Alessandro Sebastianelli$^{b}$, Maria Pia Del Rosso$^{a}$, \\ Paolo Gamba$^{c}$
\textit{and Silvia Liberata Ullo}$^{a}$\end{tabular}\thanks{
$^{1}$Corresponding author. 
\textit{Email addresses}: f.mauro$@$studenti.unisannio.it (FM), alessandro.sebastianelli$@$esa.int (AS), mpdelrosso$@$unisannio.it (MDR), paolo.gamba@unipv.it (PG), ullo$@$unisannio.it (SLU)} 
}

\address{
$^{a}$ Engineering Department, University of Sannio, Benevento, Italy \\
$^{b}$ $\phi$-lab, European Space Agency, Frascati, Italy \\
$^{c}$ Engineering Department, University of Pavia, Pavia, Italy 
}

\begin{document}
\maketitle

\begin{abstract}
The use of Synthetic Aperture Radar (SAR) has greatly advanced our capacity for comprehensive Earth monitoring, providing detailed insights into terrestrial surface use and cover regardless of weather conditions, and at any time of day or night. However, SAR imagery quality is often compromised by speckle, a granular disturbance that poses challenges in producing accurate results without suitable data processing. In this context, the present paper explores the cutting-edge application of Quantum Machine Learning (QML) in speckle filtering, harnessing quantum algorithms to address computational complexities. We introduce here QSpeckleFilter, a novel QML model for SAR speckle filtering. The proposed method compared to a previous work from the same authors %traditional filters and State-of-the-Art (SOTA) AI-based models 
showcases its superior performance in terms of Peak Signal-to-Noise Ratio (PSNR) and Structural Similarity Index Measure (SSIM) on a testing dataset, and it opens new avenues for Earth Observation (EO) applications.
\end{abstract}


\section{Introduction}
Observing the Earth through Synthetic Aperture Radar (SAR) has revolutionised our ability to monitor the planet, enabling detailed analysis of terrestrial surfaces irrespective of weather conditions. However, the acquisition of SAR images is often compromised by the phenomenon of speckle, a granular disturbance that can adversely affect the quality of results. The speckle presence can hinder accurate data interpretation and poses a significant challenge in the field of Earth Observation (EO).

The need of filtering SAR images to mitigate the effects of speckle is crucial for obtaining more reliable and action\nobreak able insights. In the quest for innovative solutions, the revolutionary potential of Quantum Machine Learning (QML) in the domain of de-speckling opens to possible investigations, aim\nobreak ing to assess the advantages that quantum algorithms can offer over classical approaches \cite{mcclean2016theory, cerezo2021variational}.
%The utilization of quantum algorithms could offer substantial advantages over classical approaches, harnessing the peculiarities of quantum computation to handle high computational complexities and enhance the efficacy of de-speckling techniques. \\
%This paper explores the context of SAR image de-speckling, examining current challenges and presenting an approach based on Quantum Machine Learning (QML). 
Through a critical review of existing methodologies and an analysis of results obtained by applying quantum algorithms to the speckle problem, we aim to highlight the innovative potentials and future prospects that Quantum Convolutional Neural Networks (QCNNs) bring in improving the quality of SAR images for EO. The amalgamation of advanced technologies such as quantum computing with EO promises to open new frontiers in understanding and monitoring our planetary environment. 
In fact, QCNNs signify a groundbreaking advancement in deep learning, merging principles of quantum computing into convolutional neural networks (CNNs). QCNNs aim to elevate visual data representation in computer vision through quantum-inspired operations.
At the core of QCNNs lie quanvolutional operations, replacing or enhancing traditional convolutions in classical CNNs. These operations leverage quantum-inspired filters capturing intricate quantum states for more expressive features. Implementation methods include quantum circuits, gates, or other quantum-inspired techniques. Utilising these filters, QCNNs can learn to extract features, exploiting the inherent quantum nature of data, leading to improved performance in various computer vision tasks \cite{wang2022development, henderson_quanvolutional_2019}.
In this paper, we present a novel QML model, the QSpeckleFilter, first pre-processing the dataset and then filtering the SAR speckle. The proposed method can overcome the others in the SOTA as will be demonstrated ahead.

\section{Background}

\begin{figure*}[!ht]
    \centering
    \includegraphics[width=2\columnwidth]{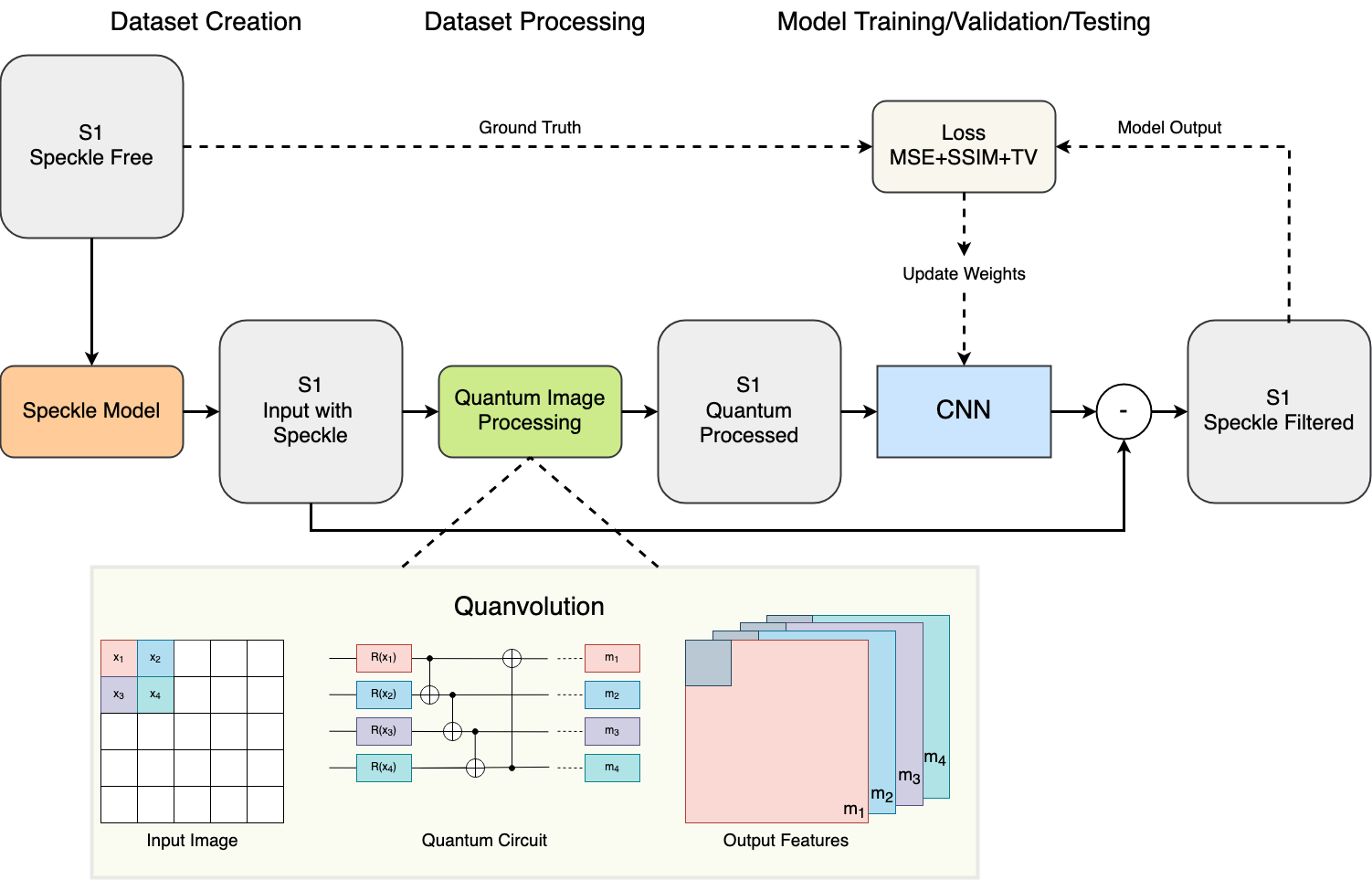}
    \caption{ The novel QML model, the QSpeckleFilter, proposed in this paper. 
    Our proposed method consists of pre-processing the input data using quantum convolution. The pre-processed data are then used to train a modified version of the speckle denoiser proposed in our previous work \cite{sebastianelli2022speckle}.}
    \label{fig:qspecklefilter}
\end{figure*}

The ongoing challenge of mitigating speckle noise in SAR images has prompted the development of various filtering methods. These methods can be broadly categorised into two main groups: \textit{Single-product speckle filtering} and \textit{Multi-temporal speckle filtering}. Within these categories, approaches can further be classified based on the underlying principles: \textit{Statistical methods} and \textit{AI-based methods}, as summarised by Table \ref{tab:categorization}.
In particular, \textbf{AI-based speckle filtering} leverages AI methods, especially CNNs. Pioneering works by Sebastianelli et al. \cite{sebastianelli2022speckle}, Chierchia et al. \cite{8128234}, Wang et al. \cite{Wang2017, wang2018generating}, Lattari et al. \cite{lattari2019deep}, Cozzolino et al. \cite{cozzolino2020nonlocal}, Dalsasso et al. \cite{dalsasso1, dalsasso2021sar2sar}, and Molini et al. \cite{molini2021speckle2void} demonstrate the application of CNNs in speckle denoising. Table \ref{tab:categorization} summarises key information about these AI-based methods, including the satellite data used for training, the SAR products, the presence of log-scaling, and the type of proposed strategy.

\begin{table}[!ht]
    \centering
    \caption{Categorization of AI-based Speckle Filters\\}\label{tab:categorization}
    \resizebox{1\columnwidth}{!}{
    \begin{tabular}{lllll}
        \toprule
        Paper & Data & SAR Products & Log-scale & Strategy\\
        \midrule
        \cite{sebastianelli2022speckle} & Sentinel-1 & GRD & no & subtractive\\
        \cite{8128234} & COSMO-SkyMED & SLC & yes & subtractive\\
        \cite{8128234} & COSMO-SkyMED & SLC & yes & subtractive\\
        \cite{Wang2017} & - & - & no & divisional\\
        \cite{wang2018generating} & RADARSAT-1 & SLC & no & divisional\\
        \cite{lattari2019deep} & Sentinel-1 & SLC & yes & subtractive\\
        & COSMO-SkyMED &  &  & \\
        \cite{cozzolino2020nonlocal} & RADARSAT & SLC & yes & non-local CNN\\
        & TerraSAR-X &  &  & \\
        & COSMO-SkyMED &  &  & \\
        \cite{dalsasso1} & TerraSAR-X & SLC & yes & subtractive\\
        & Sentinel-1 &  &  & \\
        \cite{dalsasso2021sar2sar} & Sentinel-1 & SLC & yes & subtractive\\
        \cite{molini2021speckle2void} & TerraSAR-X & SLC & no & Blind-spot CNN\\
        \bottomrule
    \end{tabular}}
\end{table}

In this paper, we opted to build upon the method proposed by Sebastianelli et al. \cite{sebastianelli2022speckle}, as it already yielded superior metric values on the test set compared to the State-of-the-Art (SOTA), QSpeckleFilter. Its implementation is shown in Figure \ref{fig:qspecklefilter} and described in the next section. 

\begin{figure*}
    \centering
    \resizebox{1.8\columnwidth}{!}{
    \begin{tabular}{lc}
    Inputs &  \includegraphics[width=2.0\columnwidth]{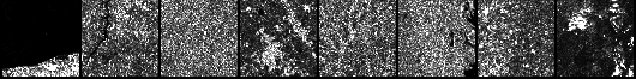}\\
    Ground Truth & \includegraphics[width=2.0\columnwidth]{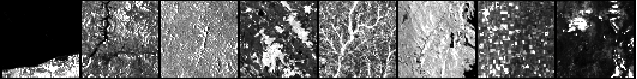}\\
    Prediction & \includegraphics[width=2.0\columnwidth]{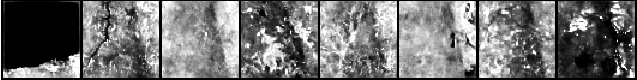}
    \end{tabular}}
    \caption{Graphical Results for the test dataset, where each column is a different sample. Inputs row refers to speckled data. Prediction row refers to QSPeckleFilter outputs.}
    \label{fig:enter-label}
\end{figure*}

\section{Data and Methods}
This section summarises first the process of creating the dataset proposed in \cite{sebastianelli2022speckle} and used for training a Neural Network suitably designed for SAR speckle filtering. The method involves obtaining a long time-series of GRD Sentinel-1 images without logarithmic scaling, from Google Earth Engine (GEE), applying temporal averaging to reduce speckle influence, and generating speckle noise using a Gamma distribution. The generated speckle is then multiplied with the ground truth (GT) to create "SAR-like" reference images, following one of the most common mathematical models for speckle also presented in \cite{sebastianelli2022speckle}. 
The synthetic GT is generated by temporal averaging intensity data, and the final input is obtained by multiplying it with the generated speckle. The dataset is created for various latitudes and longitudes, form\nobreak ing vectors of inputs and GTs. After data augmentation, %\cite{sebastianelli2022speckle}
the final dataset comprises $2637$ Sentinel-1 acquisitions without speckle and $2637$ corresponding acquisitions with speckle. Training-validation-test split factors, respectively of $75.8 \%$, $22.8 \%$ and $1.4 \%$, have been used. 
\\

\textbf{Quanvolution:}~The generated dataset was pre-processed in compliance with the new proposed method. Namely, the Quanvolutional operator (schematised in the lower part of Fig. \ref{fig:qspecklefilter}) was used in this end. The proposed Quanvolution operator is settled on the Basic Entangler circuit presented in \cite{bergholm2018pennylane} and shown in the same Figure.\\
Quanvolutions, in comparison to convolutions, yield more informative feature maps, enhancing the overall capability of the model \cite{henderson_quanvolutional_2019}. The original domain of the dataset with only one channel, namely \textit{W x H x Channels = 64 x 64 x 1}, is expanded to 64 x 64 x 9 by employing Quanvolutions. The augmentation to nine channels aids the subsequent neural network in more effectively extracting noise from the data.

We opted for a 9-qubits configuration to employ a 3x3 kernel, representing a good balance since the pre-processing time, given current technological constraints, has remained within a reasonable range ($\le 1$ day). As regards the choice of the circuit, the rationale behind this lies in the simplicity of the circuit's topology, which has consistently demonstrated effective performance across various scenarios, as evidenced by Incudini et al. \cite{incudini2023resource}. In future work, we want to explore alternative configurations to this circuit, by investigation and utilization of optimization methods, as proposed by Sebastianelli et al. in  \cite{sebastianelli2023quantum}.  {\color{blue} The new pre-processed dataset and the implemented code will be made available to the public upon paper publication.}

\textbf{Speckle Denoiser:}~After pre-processing the dataset with quanvolution, we have exploited and properly modified the architecture proposed in \cite{sebastianelli2022speckle}, to work with the new dataset.\\
Our methodology uses the quanvoluted feature maps to estimate the speckle noise present in each image. The extracted speckle is subsequently subtracted from the Sentinel-1 image through a \textit{skip connection} and a \textit{subtraction layer}, rendering the model residual. The skip connection acts as a detour for CNN. The subtraction layer executes mathematical subtraction between the input data transmitted through the skip connection and the CNN output. \\
In contrast to methods converting the multiplicative speckle model into an additive one or those employing divisional layers, the proposed approach embraces a residual strategy centered on subtraction layers. It mitigates potential computational errors associated with divisional layers, and it avoids logarithmic scaling of data to preserve statistical characteristics, by utilizing GEE data in the FLOAT format. \\
The residual noise, expressed as the difference between the noisy and clean images, undergoes optimization during training to learn the mapping function. \\
The model minimizes a customized loss function given by equation \ref{eqn:loss}, proposed by Sebastianelli et al. in \cite{sebastianelli2022speckle}, encompassing terms for pixel-wise Euclidean distance, preservation of structural similarity, and Total Variation Loss to promote smoother results. 
\begin{equation}
    \centering
    \begin{split}
    L&(x ,x^*) = \alpha  \frac{1}{WH} \sum^W_{w=1}\sum_{h=1}^H [x_{w,h} - x^*_{w,h}]^2 + \\
     &+ \beta \frac{(2\mu_{x}\mu_{x^*}+c_1)(2\sigma_{xx^*}+c_2)}{(\mu^2_{x}+\mu^2_{x^*}+c_1)(\sigma^2_{x}+\sigma^2_{x^*}+c_2)} +\\
     &+ \gamma \sum_{w=1}^{W}\sum_{h=1}^H \sqrt{(x_{w+1,h}-x_{w,h})^2 + (x_{w,h+1}-x_{w,h})^2}
    \end{split}
    \label{eqn:loss}
\end{equation}
\section{Results}
From a qualitative perspective, the results in Fig. \ref{fig:enter-label} exhibit a favorable outcome, displaying a smoothing effect on the filtered images overcoming the findings already highlighted in  Sebastianelli et Al. \cite{sebastianelli2022speckle}. The quality of the results is also substantiated by the quantitative values presented in Table \ref{tab:res}.
The application of quanvolution in data processing has proven to be a transformative enhancement, leading to a good improvement in performance metrics such as PSNR and SSIM, which increased respectively by 13\% and 8\%. As evident from the results, our approach surpasses the metrics of our previous work, %\cite{sebastianelli2022speckle}, 
reaffirming the potential of quanvolution in image processing. The key differentiator, from our previous work,%\cite{sebastianelli2022speckle}, 
contributing to the notable enhancement, mainly consists of preprocessing the dataset throughout the quanvolution operator. So, we achieved better results without altering the complexity of the network. This underscores the efficacy of quanvolution as a powerful technique, elevating the overall capabilities of our model compared to existing SOTA algorithms, whose comparison and analysis were already carried on in \cite{sebastianelli2022speckle}.

\begin{table}[!ht]
    \centering
    \caption{Proposed model's average scores on the testing dataset: (a) GT, (b) Input with speckle, (c) Sebastianelli et Al. \cite{sebastianelli2022speckle}
    and (d) Proposed \textbf{QSpeckleFilter.}\\}\label{tab:res}
    \begin{tabular}{lcc}
    \toprule
    Model & PSNR $\uparrow$ & SSIM $\uparrow$ \\
    \midrule
    (a) Ground Truth & $+\infty$ & 1.0 \\
    (b) Speckled & 15.70 & 0.58 \\
    \midrule
    %(c) Lee & 16.64 & 0.61 \\
    %(d) Lee Enanched & 16.46 & 0.55 \\
    %(e) Kuan & 16.78 & 0.61 \\
    %(f) Frost & 16.93 & 0.61 \\
    %(g) Mean & 15.93 & 0.50 \\
    %(h) Median & 15.39 & 0.48 \\
    %(i) Fastnl & 15.67 & 0.58 \\
    %(j) Bilateral & 17.29 & 0.62 \\
    %(k) SAR-BM3D & 15.67 & 0.58 \\
    %\midrule
    (c) Sebastianelli et Al. \cite{sebastianelli2022speckle} & 19.21 & 0.75 \\
    (d) \textbf{QSPeckleFilter} & 21.72 & 0.81 \\
    \bottomrule
    \end{tabular}
\end{table}
  
\section{Discussion and Conclusions}
Our study introduces QSpeckleFilter that combines quantum convolution with a speckle denoiser for SAR images. Results demonstrate superior performance in PSNR and SSIM compared to our previous method. The dataset pre-processed with quanvolutions will be publicly available after paper acceptance. This innovative approach showcases the potential of quantum-inspired operations in image filtering. The evaluation highlights the effectiveness of our method in enhancing SAR image quality.
In conclusion, this study contributes to the intersection of QML and SAR image processing, offering a promising avenue for EO applications. The public release of the pre-processed dataset encourages further exploration and collaboration in this emerging field.
\vspace{-0.4cm}
\balance
\bibliographystyle{IEEEtran}
\bibliography{strings,refs}

\end{document}